\documentclass[prl,nofootinbib,twocolumn,
amsmath,amssymb,superscriptaddress]{revtex4}
\usepackage{amsmath}
\usepackage{pstricks}

\newcommand{\eqn}[1]{eq.~(\ref{#1})}

\newcommand{\beq}{\begin{equation}}
\newcommand{\eeq}{\end{equation}}
\newcommand{\beqa}{\begin{eqnarray}}
\newcommand{\eeqa}{\end{eqnarray}}

\def\lsim{\ \rlap{\raise 3pt \hbox{$<$}}{\lower 3pt \hbox{$\sim$}}\ }
\def\gsim{\ \rlap{\raise 3pt \hbox{$>$}}{\lower 3pt \hbox{$\sim$}}\ }

\begin{document} 

\title{
Lepton Flavor Equilibration and Leptogenesis
}

\author{D. Aristizabal Sierra} \email{diego.aristizabal@lnf.infn.it}
    \affiliation{INFN, Laboratori Nazionali di Frascati, C.P. 13,
      100044 Frascati, Italy}

\author{Marta Losada} \email{malosada@uan.edu.co}
    \affiliation{Centro de Investigaciones, Universidad Antonio Nari\~{n}o, \\
Cra 3 Este No 47A-15 Bloque 4, Bogot\'{a}, Colombia}

\author{Enrico Nardi} \email{enrico.nardi@lnf.infn.it}
    \affiliation{INFN, Laboratori Nazionali di Frascati, C.P. 13,
      100044 Frascati, Italy}
    \affiliation{Instituto de F\'{i}sica, Universidad de Antioquia,
    A.A.1226, Medell\'{i}n, Colombia}




\vspace{2mm}


\begin{abstract}
  We study the role played in leptogenesis by the equilibration of lepton
  flavors, as could be induced in supersymmetric models by off diagonal soft
  breaking masses for the scalar lepton doublets $\tilde m_{\alpha\beta}$, or
  more generically by new sources of lepton flavor violation.  We show that if
  $\tilde m_{\alpha\beta}\gsim 1\,$GeV and leptogenesis occurs below $\sim
  100\,$TeV, dynamical flavor effects are irrelevant and leptogenesis is
  correctly described by a one-flavor Boltzmann equation.  We also discuss
  spectator effects in low scale leptogenesis by taking into account various
  chemical equilibrium conditions enforced by the reactions that are in
  thermal equilibrium.  We write down the Boltzmann equation for low scale
  supersymmetric leptogenesis that includes flavor and spectator effects in
  the presence of lepton flavor equilibration, and we show how it reduces to a
  particularly simple form.
\end{abstract}
%

\maketitle



\baselineskip 16pt

\section{Introduction}
 \label{sec:intro}
\vspace{-3mm}
Leptogenesis~\cite{fu86} is a theoretical mechanism that can explain the
observed matter-antimatter asymmetry of the Universe.  An initial lepton
asymmetry is generated in the out-of-equilibrium decays of heavy singlet
Majorana neutrinos, and is then partially converted in a baryon asymmetry by
anomalous sphaleron interactions~\cite{Kuzmin:1985mm}.  Heavy Majorana singlet
neutrinos are also a fundamental ingredient of the seesaw model~\cite{seesaw},
that provides an elegant explanation of the suppression of  neutrino masses
with respect to all other Standard Model (SM) mass scales.  Leptogenesis can
be quantitatively successful without any fine-tuning of the seesaw parameters,
and it is an intriguing `coincidence' that a neutrino mass scale of the order
of the atmospheric neutrino mass square difference is optimal for yielding the
correct value of the baryon asymmetry.  The possibility of giving an
explanation of two apparently unrelated experimental facts (neutrino masses
and the baryon asymmetry) within a single framework, makes the idea that
baryogenesis occurred through leptogenesis  very attractive.

The crucial role played by lepton flavor effects in leptogenesis was fully
realized only a few years ago \cite{aba06a,Nardi:2006fx,aba06b} (see
\cite{ba00,en03-fu05} for earlier studies of flavor effects in leptogenesis,
and \cite{Davidson:2008bu,review} for recent reviews).  Most extensions of the
SM, and most noticeably among these, the Supersymmetric Standard Model (SSM),
include new sources of Lepton Flavor Violation (LFV).  The purpose of this
letter is to show that if leptogenesis occurs at temperatures when these new
sources mediate reactions that are in chemical equilibrium, then there are no
flavor effects in leptogenesis.\footnote{In this paper, we refer to ``flavor
  effects'' in the restricted sense of the dynamical effects that arise when
  the flavored CP asymmetries $\epsilon_\alpha$ are not proportional to the
  respective branching fractions for $N\to \ell_\alpha$ decays. In the (fine
  tuned) case when an exact proportionality exists, the final baryon asymmetry
  is simply enhanced by a factor corresponding to the number of lepton flavors
  that are in thermal equilibrium~\cite{Nardi:2006fx}. 
 This result holds independently of the particular flavor
 configuration, it is unrelated to flavor dynamics, and is a simple
 consequence of fermion family replication. Thus, we include this
 effect in the general class of effects that, like the typical
 spectator effects, are unrelated to the dynamics of lepton flavors
 and of lepton flavor number violation.
}  In particular, this is likely to happen in
soft leptogenesis \cite{softlepto,D'Ambrosio:2003wy,Grossman:2004dz} that is a
natural mechanism for low scale leptogenesis.  In contrast, 
We include in our analysis also the effects of spectator processes
\cite{bu01,na05} that should be taken into account for a correct estimate of
the final baryon asymmetry.
  
\section{Lepton Flavor Equilibration 
}
 \label{sec:LFE}
\vspace{-3mm}
With Lepton Flavor Equilibration (LFE) we refer to the effect of reactions
that would bring the different lepton doublets $\ell_\alpha $
($\alpha=e,\mu,\tau$) into chemical equilibrium.  We will use as a general and
most interesting example the SSM where, in the basis in which the charged
lepton Yukawa couplings are diagonal, a source of LFV from soft supersymmetry
breaking masses is generally present:
\begin{equation}
\label{eq:soft}
{\cal L}_{{\rm soft}}\ \supset \ \tilde m^2_{\alpha\beta}\tilde \ell^\dagger_\alpha
\tilde \ell_\beta.
\end{equation}
Here $\tilde \ell_\alpha$  are
the superpartners of the $SU(2)$ lepton doublets.  
The terms in \eqn{eq:soft} affect the flavor composition of the mass
eigenstates, and as a result  the $\ell_\alpha \tilde \ell_\alpha^{({\rm int})} \tilde G$ vertex for
the sleptons gauge eigenstates (where $\tilde G= \tilde W_a, \tilde B$
represent a $SU(2)$ or $U(1)$ gaugino) involve a unitary rotation to the slepton mass
eigenstates:
\begin{equation}
\label{eq:rotation}
 \tilde \ell_\alpha^{({\rm int})} = R_{\alpha\beta} \tilde \ell_\beta, \qquad  
R_{\alpha\beta} \sim \delta_{\alpha\beta} + 
{\cal O}(\frac{\tilde m^2_{\alpha\beta}}{h^2_\alpha T^2}),
\end{equation}
where $h_\alpha > h_\beta$ is the relevant charged lepton Yukawa coupling that
determines at leading order the (thermal) mass splittings of the sleptons.

The term in \eqn{eq:soft} can induce fast LFV reactions,
namely gaugino mediated $t$-channel processes $\ell_\alpha P \leftrightarrow
\tilde \ell_\beta \tilde P$, $\ell_\alpha \tilde P \leftrightarrow \tilde \ell_\beta \ 
P$, and $s$-channel processes $\ell_\alpha \tilde \ell_\beta \leftrightarrow P
\tilde P$, with $P=\ell,Q,\phi_u,\phi_d$, where $Q$ and $\phi_{u,d}$ 
are the quark and Higgs doublets 
(for bino-reactions also the quark and leptons $SU(2)$-singlets
$u,d,e$ contribute).  For example, for the 
reduced cross section corresponding to $SU(2)$ $t$-channel
reactions
we obtain: 
\begin{eqnarray}
\nonumber
\hat\sigma_t (s) &=& 
R_{\alpha\beta} R^*_{\alpha\beta} \frac{g^4}{16\pi} \Pi_g \times \\
%
\label{eq:reduced}
&&
\left[\left(2\frac{m_{\tilde W}^2}{s}+1\right) \log\left(\frac{s+m_{\tilde
        W}^2}{m_{\tilde W}^2}\right)-2\right] 
\end{eqnarray}
where $\Pi_g\equiv \left(\sum_{\beta\neq\alpha } g_{L_\alpha} g_{\tilde
    L_\beta}\right) \times \sum_P g_P g_{\tilde P}$ where $g_L,g_P,\dots $
count the number of degrees of freedom (spin, isospin, color and flavor) of
the corresponding particles, and $m_{\tilde W}^2 =(9/2)\, g^2_2 T^2$ is the
$W$-ino thermal mass.  Because of the several possible reactions and of the
large number of degrees of freedom, $\Pi_g$ is a large number $\sim 10^3$.
From \eqn{eq:reduced}  and using  eq.~(\ref{eq:rotation}) (for
$\alpha\neq\beta$ that is the case of interest) we can estimate
the thermally averaged reaction density for $t$-channel 
$\alpha\leftrightarrow\beta$ transitions as: 
%
%
\begin{eqnarray}
\nonumber
\gamma_{t, \alpha\leftrightarrow\beta} 
&=& \frac{T}{64 \pi^4}
\int ds \sqrt{s} K_1\left(\frac{\sqrt{s}}{T}\right)\hat\sigma_t(s) \\
&&  \sim \left(\frac{g^2\tilde m_{\alpha\beta}^2}{h^2_\alpha T^2}\right)^2
\frac{\Pi_g T^4}{2^9\pi^5}. 
\end{eqnarray}
Summing  $s$-channel reactions and normalizing to the relativistic abundance
of  leptons we obtain the LFV rate
\begin{equation}
\label{eq:feq}
\Gamma_{LFV} \sim
\frac{\gamma_{\alpha\leftrightarrow\beta}}{ T^3/\pi^2} \sim 
10^6\, \frac{\tilde m_{\alpha\beta}^4}{T^3}.
\end{equation}
When this rate is faster than the Universe expansion
$\Gamma_H \sim 25\, T^2/M_P$ (with $M_P$ the Planck mass), 
asymmetries eventually present in the different lepton doublets 
equilibrate, meaning that their chemical potentials (that here and in the
following are denoted with the same symbol than the corresponding particle) 
become equal: $\ell_\alpha  =\ell$. According to 
\eqn{eq:feq}, this occurs roughly for 
\begin{equation}
  \label{eq:Teq}
  T\lsim 100\, \left(\frac{\tilde m_{\alpha\beta}}{1\,{\rm GeV}}\right)^{4/5}\, {\rm TeV}.
\end{equation}
We see that even for moderate values of the off-diagonal soft breaking scalar
masses, LFE is likely to be a generic feature in low scale
supersymmetric leptogenesis.

\section{Equilibrium Conditions}
 \label{sec:equilibrium}
\vspace{-3mm}
To see what are the consequences of LFE on leptogenesis, we need to write the
corresponding Boltzmann Equations (BE) taking into account all the chemical
equilibrium conditions imposed by reactions that, at the specific temperature
considered, are faster than the Universe expansion.  Here we will concentrate
on the temperature range 1\,TeV$\,\lsim T\lsim\,$ 100\,TeV, that is well above
the electroweak phase transition, but low enough so that in the SSM, LFE
equilibration is likely to occur.
In principle there are as many chemical potentials as there are
particles in the  thermal bath. However, a first  set of conditions
that are generally realized  in the temperature range we are interested in 
allows to drastically reduce this number:

\begin{enumerate}
  
\item Since we will work at scales much higher than $M_W$, where total isospin
  $I_3$, hypercharge (and color) must be zero, gauge fields have vanishing
  chemical potential $W=B=g=0$~\cite{ha90}. This also implies that all the
  particles belonging to the same $SU(3)\times SU(2)\times U(1)$ multiplet
  have the same chemical potential, that is,
  $\psi(I_3=+\frac{1}{2})=\psi(I_3=-\frac{1}{2})$ for weak isospin, and
  similarly for color.
  
\item Chemical potentials for the three gauginos $\tilde W=\tilde B=\tilde g$
  are driven to zero once the supersymmetry breaking effects related to the
  Majorana soft gaugino masses $m_{1/2}$ attain chemical
  equilibrium~\cite{susy-equilibrium}, that is when $m_{1/2}^2/T\gsim \Gamma_H
  $.  Note that since these rates are quadratic in $m_{1/2}$ while the
  corresponding rates for LFE go as $\tilde m^4_{\alpha\beta}$ in studying LFE
  it is certainly reasonable to assume $\tilde W=\tilde B=\tilde g=0$.  In turn,
  the vanishing of the gaugino chemical potentials implies that particles
  within the same supermultiplet have the same chemical potential $\tilde
  \psi=\psi$.
  
\item Similarly to what happens for the gauginos, the Higgsino mass term
  $\mu_{\phi_{u,d}} \phi_u\phi_d$ ensures that $\phi_d+ \phi_u=0$. We then
  denote $\phi_u=- \phi_d = \phi$.
  
\item Because of generation-mixing interactions, we take
  generation-independent quark potentials $Q_i=Q$, $u_i=u$ and $d_i=d$.    
 Fast LFV interactions yield  ($\alpha\neq \beta$)
\begin{equation}
\ell_\alpha + P = \tilde\ell_\beta + \tilde P = \ell_\beta + P,  
\end{equation}
where $P$ represents, for example, any one of the $\ell,\,Q,\,\phi_{u,d}$
$SU(2)$ doublets, and the second equality follows from condition 2.
This yields $\ell_\alpha = \ell_\beta=\ell$, that is the chemical
potential of the lepton doublets are also generation independent.

\item Given that $\ell_\mu=\ell_\tau$, the Yukawa couplings interactions for
  the right handed $\mu$ and $\tau$ leptons, that are both in equilibrium
  below $ (1 + \tan^2\beta)\times10^9\,$GeV, yield $\tau=\mu$.  As regards the
  right handed electron, its Yukawa interaction attains equilibrium below
  $T_e\sim 20-100\,$TeV~\cite{eR-equilibrium}, depending also on the value of
  $\tan\beta$.\footnote{Reactions mediated by the small up-quark Yukawa
    coupling $Y_u$ will be in equilibrium at temperatures below $T_u \sim
    (Y_u/Y_e)^2 T_e \approx (m_u^2/m_e^2\tan^2\beta)\, T_e$. Thus, if
    $\tan\beta < m_u/m_e \approx {\cal O}(10)$, reactions mediated by $Y_u$
    will always be in equilibrium when the electron Yukawa interactions are in
    equilibrium.  If $\tan\beta \gg m_u/m_e$, the up-quark Yukawa interactions
    can remain out-of-equilibrium in most of the temperature regime we are
    considering. This would yield slightly different numerical results.} If
  also the off diagonal mass terms for the right handed sleptons $\tilde
  m^R_{e\alpha}$ ($\alpha=\mu,\tau$) are particularly suppressed, then the
  right handed (s)electrons would remain out of chemical equilibrium in an
  interesting range above $T_e$. We thus leave open the possibility that
  $\mu\neq e=0$.

\end{enumerate}

Because of the previous conditions, we are left with 
six independent chemical potentials:
$Q,\,u,\,d,\,\ell,\,\mu$ and $\phi$.
They should satisfy the following additional conditions:

Yukawa couplings equilibration:
\begin{eqnarray}
  \label{eq:Yukawau}
Q-u+\phi_u&=&0, \\
  \label{eq:Yukawad}
Q-d+\phi_d&=&0, \\
  \label{eq:Yukawal}
\ell-\mu+\phi_d&=&0.  
\end{eqnarray}
Electroweak sphalerons equilibrium (QCD sphalerons equilibrium  do not impose 
further constraints~\cite{na05}):
\begin{equation}
  \label{eq:sphalerons}
3Q+\ell=0.
\end{equation}
Hypercharge neutrality:
\begin{equation}
  \label{eq:hypercharge}
3(Q+2u-d-\ell) -2\mu -e +\phi_u-\phi_d=0,  
\end{equation}
where, depending on the temperature range,
$e=\mu$ or $e=0$. 
These are five conditions, and therefore all the chemical potentials can be 
expressed in terms of just one, that we choose to be $\ell$. 
The solution for $e=\mu$ reads: 
\begin{equation}
\label{eq:solutions}
Q= -\frac{\ell}{3}; \
u=  \frac{5\ell}{21};  \
d= -\frac{19\ell}{21}; \
\mu = \frac{3\ell}{7};  \
\phi= \frac{4\ell}{7},
\end{equation}
with minor numerical changes in $u,\,d,\,\mu,\,\phi$ 
when $e=0$. 
%
We can now use these equations to express $\ell$ in terms of the asymmetry in
the $B-L$ charge, that is the relevant quantity for writing the BE in our
temperature regime since it is not violated by EW sphalerons.  We denote the
number density asymmetry for a particle $p$ normalized to the entropy density
$s$ as $Y_{\Delta p}={(n_p-n_{\bar p})}/{s}$, and  $Y_{\Delta p}$ normalized 
to the equilibrium density $Y_p^{{\rm eq}}$ is denoted as 
$y_{\Delta p} = Y_{\Delta p}/Y_p^{{\rm eq}}$.
Let us also remember that, because of boson (B)/fermion (F) statistics,
the relation between chemical potentials and particle density asymmetries 
reads $y_{\Delta B}/y_{\Delta F } = 2
\mu_{B}/\mu_{F}$.  
Assuming that all effects of particle masses can be neglected~\cite{Rubakov:1996vz}, 
and depending if the right-handed electron is 
or is not in chemical equilibrium, 
then we  have:
\begin{eqnarray}
  \label{eq:B-L}
\frac{Y_{\Delta_{B-L}}^{(e=\mu)}}{Y_{\Delta\ell}}\!\!\! &=&\!\! \frac{9}{\ell} \left[2Q+u+d-(2\ell+\mu)
\phantom{\frac{1}{1}}\!\!\!\!
\right]\! =-\frac{237}{7}\,,  \\
  \label{eq:B-Lnoeq}
\frac{Y_{\Delta_{B-L}}^{(e=0)}}{Y_{\Delta\ell}}\!\!\! &=&\!\! 
\frac{9}{\ell} \left[2Q+u+d-\left(2\ell+\frac{2}{3}\mu\right)\!\right] \!
=-\frac{426}{13}\,.\ \ \quad
\end{eqnarray}

\section{The Boltzmann Equations}
 \label{sec:BE}
\vspace{-3mm}
When accounting for flavor effects is a mandatory condition ($T\lsim
10^{12}\,$GeV~\cite{aba06a,Nardi:2006fx,ba00}) one should write a set of three
BE, one for each of the conserved flavor charges $\Delta_\alpha=B/3-L_\alpha$.
To illustrate the importance of LFE, here we write simplified BE including
only decays and inverse decays. Also, we write just the BE for the heavy
Majorana neutrinos, the effects of LFE on the equations for the sneutrinos are
completely analogous.  The BE for the evolution of the heavy neutrino density
as usual reads: $-\dot Y_{N} = (y_N-1) \gamma_D$ where $\gamma_D$ is the total
$N$ decay rate into $\ell,\phi_u$ and their superpartners, $y_N =
{Y_{N}}/{Y^{{\rm eq}}_N}$ and the time derivative is $\dot Y_{N}= sHz\,d
Y_{N}/dz$, with $z=M_N/T$.  By taking equal decay rates into particles and
superpartners $\gamma_{D_\ell}\simeq \gamma_{D_{\tilde\ell}} \simeq
\gamma_D/2$ we can write the evolution equation for the flavor charges as:

\begin{eqnarray}
\nonumber
-\dot Y_{\Delta_\alpha} &=&  (y_N-1) \epsilon_\alpha \gamma_D
\\
&& \!\!\!\!\!\!\!\!\!\! \!\!\!\!\!
-  
 \frac{1}{2} 
\left(y_{\Delta\ell}+y_{\Delta\phi} 
 + y_{\Delta\tilde\ell}+y_{\Delta\tilde\phi}\right)\, 
B_\alpha \left(\frac{1}{2} \gamma_{D}\right)  \nonumber \\
 \label{eq:BE2}
&& \!\!\!\!\!\!\!\!\!\! \!\!\!\!\!
 = (y_N-1) \epsilon_\alpha \gamma_D
-  \frac{1}{2} \left(\frac{33}{7}\, y_{\Delta\ell}\right)\, 
B_\alpha \left(\frac{1}{2} \gamma_{D}\right)\,,  
\end{eqnarray}
where $\epsilon_\alpha$ and $B_\alpha$ are respectively the CP asymmetry and
the branching fraction for the decay $N \to \ell_\alpha+\tilde\ell_\alpha$.
In the last line we have used $y_{\Delta\phi}= \frac{8}{7}y_{\Delta\ell}$,
$y_{\Delta\tilde \phi}= \frac{4}{7}y_{\Delta\ell}$ and
$y_{\Delta\tilde\ell}=2y_{\Delta\ell}$.  Now in order to integrate
\eqn{eq:BE2}, one should express the asymmetry density $y_{\Delta\ell}$ that
is weighting the strength of the washouts, in terms of the charge densities
$y_{\Delta_\alpha}$.  The important thing to notice at this point is that,
regardless of the details of the resulting expression, as a consequence of LFE
the washout weights do not carry any flavor index.  We can thus readily sum up
the three flavored equations and obtain
\begin{equation}
  \label{eq:BE2a}
-\dot Y_{\Delta_{B-L}} =  (y_N-1) \epsilon\, \gamma_D
+  \frac{1}{4} \left(\frac{11}{79}\right)\,  y_{\Delta_{B-L}} \gamma_{D},   
\end{equation}
where $\epsilon = \sum_\alpha \epsilon_\alpha$ and \eqn{eq:B-L} has been used.
For the case in which the right handed electron is out of equilibrium, 
\eqn{eq:B-Lnoeq} together with $y_{\Delta\phi}= 2 y_{\Delta\tilde
  \phi}=\frac{14}{13}y_{\Delta\ell}$ should be used instead, and the washout
numerical coefficient changes slightly: $\frac{11}{79}\to \frac{10}{71}$.


%
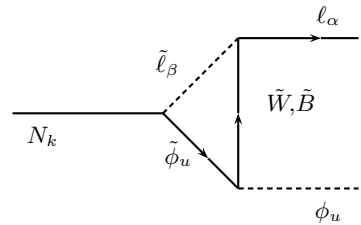
\begin{figure}[t]
\begin{center}
  \begin{pspicture}(-7.0,2.9)(2.7,5.5)
 \psset{xunit=2.0cm}
 \psset{yunit=2.0cm}
    \psline(-2.5,2)(-1.5,2)
    \psline[linestyle=dashed,dash=2pt 1.5pt](-1.5,2)(-1.0,2.5)
    \psline{->}(-1.5,2)(-1.2,1.7)
    \psline(-1.2,1.7)(-1.0,1.5)
    \psline{->}(-1.0,1.5)(-1.0,2.0)
    \psline(-1.0,2.0)(-1.0,2.5)
    \psline{->}(-1.0,2.5)(-0.45,2.5)
    \psline(-0.45,2.5)(-0.2,2.5)
    \psline[linestyle=dashed,dash=2pt 1.5pt](-1.0,1.5)(-0.2,1.5)
    \uput[d](-2.3,2){{$N_k$}}
    \uput[u](-1.47,2.15){{$\tilde \ell_\beta$}}
    \uput[u](-0.4,2.5){{$\ell_\alpha$}}
    \uput[u](-0.65,1.9){{$\tilde W\!,\!\tilde B$}}
    \uput[d](-1.4,1.9){{$\tilde\phi_u$}}
    \uput[d](-0.4,1.5){{$\phi_u$}}
  \end{pspicture}
\end{center}
\caption{Vertex diagram generating  lepton flavor violating CP asymmetries  
  in the decays of the heavy Majorana neutrinos $N_k \to \ell_\alpha \phi_u$.
  Similar diagrams appear for $N_k \to \tilde \ell_\alpha \tilde \phi_u$ and in
  the decays of the neutrinos superpartners $\tilde N_k \to \ell_\alpha
  \tilde \phi_u, \tilde \ell_\alpha \phi_u$.}
\label{fig:diagram1}
\end{figure}

A few remarks are now in order: 

\begin{enumerate}
  
\item LFE allows to recast the BE equation in just one equation for $B-L$,
  thus killing effectively  flavor effects.  Recalling that in the
  temperature regime $T\lsim 10^9\,$GeV also all effects of the heavier
  neutrinos $N_{2,3}$ are absent~\cite{Engelhard:2006yg}, we can conclude that
  the one-flavor BE \eqn{eq:BE2a} gives a complete description of the dynamics
  of leptogenesis.  Note that such a drastic simplification depends just on the
  hypothesis of flavor equilibration, and
  it applies in particular to soft leptogenesis as long as $\tilde
  m_{\alpha\beta} \gsim 1\,$GeV and $M_N \lsim 100\,$TeV.  In case $\tilde
  m_{\alpha\beta}$ is diagonal to an extremely good approximation, then flavor
  effects can survive,\footnote{For flavor effects to kick back in it is in
    fact sufficient to have e.g., $\tilde m_{e\alpha}\approx 0$ for
    $\alpha=\mu,\tau$.  Below $T_e$ also $\tilde m^R_{e\alpha}\approx 0$ is
    required.} and  can have important consequences also 
   for soft leptogenesis~\cite{softflavor}.

\item Purely Flavored Leptogenesis (PFL) models~\cite{PFL}, that are defined by
  the condition $\sum_\alpha \epsilon_\alpha=0$, are unable to generate a
  baryon asymmetry in the presence of LFE since the source term in the BE
  \eqn{eq:BE2a} vanishes.\footnote{ In the PFL model studied in~\cite{PFL} LFE
    does not occur, because LFV is coupled to heavy messengers states with
    mass several times larger than $M_N$, and thus it is strongly suppressed.}
  In this respect, it is interesting to note that the same LFV soft masses
  responsible for LFE also generate lepton flavor violating CP asymmetries
  through interference of the tree level decay amplitude with the loop diagram
  depicted in fig.~\ref{fig:diagram1}.  The  CP asymmetry in
  the decay of the $N_k$ seesaw neutrino into SM particles induced by this
  diagram reads:
\begin{equation}
\label{eq:CP}
\phantom{aaa} \epsilon^{N_k}_{\ell_\alpha\phi_u} 
=-\frac{3 g^2+{g'}^2}{8\pi\,{\left(\lambda^\dagger\lambda\right)_{kk}}}\, 
\sum_\beta {\rm Im}  \left(\frac{\lambda^*_{\beta
      k}\tilde m^2_{\beta\alpha}\lambda_{\alpha k}}{M^2_{N_k}}\right), 
\end{equation}
and similar expressions hold for the other flavored CP asymmetries
$\epsilon^{N_k}_{\tilde \ell_\alpha\tilde \phi}$, $\epsilon^{\tilde
  N_k}_{\tilde \ell_\alpha \phi}$, and $\epsilon^{\tilde N_k}_{
  \ell_\alpha\tilde \phi}$.  From~\eqn{eq:CP} it is readily seen that
$\sum_\alpha \epsilon^{N_k}_{\ell_\alpha\phi_u}=0$.  In the absence of LFE,
this condition by itself would not impede to generate a lepton
asymmetry~\cite{Nardi:2006fx,PFL} and, rather interestingly, the decoupling of
the CP asymmetries (enhanced as $\sim g^2$) from the washouts ($\propto
\lambda^2$) would have yielded a new mechanism for low scale leptogenesis and
for avoiding the gravitino problem.  However, to obtain sufficiently large CP
asymmetries one must require $\tilde m^2_{\alpha\beta}/M^2_{N}\gsim 10^{-6}$,
and this unavoidably implies LFE. We can thus conclude that the flavored CP
asymmetries~\eqn{eq:CP} are irrelevant to leptogenesis.

\end{enumerate}

\section{Conclusions}
 \label{sec:conclusions}
\vspace{-3mm}
Flavor effects can produce large enhancements of the baryon asymmetry yield of
leptogenesis. If the CP asymmetry in one lepton flavor is particularly large
(and note that it could even be larger than the total lepton CP asymmetry) and at the same
time the associated washouts are suppressed by a small branching fraction for
$N$ decays to that flavor, the baryon asymmetry will be sizeably larger
than what would be obtained in a one-flavor approximation  based on the
total lepton CP asymmetry and on the total washout rates.  

The effectiveness of flavor effects relies, however, on the condition that the
dynamics of the different lepton flavors remains sufficiently decoupled during
the leptogenesis era.  In this paper we have shown that when LFV interactions
are sufficiently fast to equilibrate the asymmetries in the different lepton
doublets, flavor effects disappear. In this situation, the
dynamics of leptogenesis can be again described correctly by means of a single
BE for the evolution of B-L.

It is well known that in the SM+seesaw, leptogenesis remains a high energy
mechanism, and that to lower the leptogenesis scale, say  below  $\sim
10^8\,$GeV, physics beyond the SM+seesaw is required.  However, new sources of
LFV are a quite common features in models for new physics, and thus when
leptogenesis is embedded within new physics models, it is important to verify
if flavor effects are relevant or not in that particular realization.

In this paper, we have illustrated this point by studying the SSM+seesaw, that
probably is the most interesting example.  The SSM allows to produce a baryon
asymmetry at a scale as low as a few TeVs, through the mechanism of soft
leptogenesis.  However, the SSM also includes new sources of LFV from the
supersymmetry soft breaking sector, and we have shown that if the off-diagonal
soft masses for the scalar lepton doublets are larger than about $ 1\,$GeV,
then there are no flavor effects in soft leptogenesis, as long as it occurs
below roughly $T\sim 100\,$TeV.  We conclude that in the presence of LFE, a
simple and qualitatively correct description of leptogenesis can be given in
terms of just one equation for the evolution of the lepton asymmetry. However,
for a numerically accurate estimate, an equation for the evolution of $B-L$
(like eq.~(\ref{eq:BE2a})) that takes into account electroweak sphalerons
equilibrium as well as the effects of other spectator processes, is required.



 \section*{Acknowledgments}
We thank S. Davidson for illuminating conversations during the early stages of
this work, and M.C. Gonzalez-Garcia for several remarks on the preliminary
draft. M.L. acknowledges the hospitality of IPN-Lyon, LNF-Frascati and IFT-Madrid during the completion of this work.  M.L. was supported by the Ecos-Nord program. The work of M.L. and E.N. is supported in part by Colciencias under
contract number 1115-333-18739.


\begin{thebibliography}{99}

\bigskip

\bibitem{fu86} M.~Fukugita and T.~Yanagida,
  Phys.\ Lett.\ B {\bf 174}, 45 (1986). 


\bibitem{Kuzmin:1985mm}
  V.~A.~Kuzmin, V.~A.~Rubakov and M.~E.~Shaposhnikov,
  Phys.\ Lett.\ B {\bf 155}, 36 (1985).

\bibitem{seesaw} P. Minkowski, 
{\it Phys. Lett.} B {\bf 67} 421 (1977); 
T. Yanagida, in {\it Proc. of Workshop on Unified Theory and Baryon
number in the Universe}, eds. O. Sawada and A. Sugamoto, KEK, Tsukuba, (1979) p.95;
M. Gell-Mann, P. Ramond and R. Slansky,  in {\it Supergravity}, eds P. 
van Niewenhuizen and D. Z. Freedman (North Holland, Amsterdam 1980) p.315;
P. Ramond, {\it  Sanibel talk}, retroprinted as hep-ph/9809459;
S. L. Glashow, in {\it Quarks and Leptons}, Carg\`ese lectures, eds M. L\'evy,
(Plenum, 1980, New York) p. 707;
R. N. Mohapatra and G. Senjanovi\'c, {\it Phys. Rev. Lett.} {\bf 44}, 912 (1980).

 \bibitem{aba06a}
  A.~Abada, S.~Davidson, F.~X.~Josse-Michaux, M.~Losada and A.~Riotto,
  JCAP {\bf 0604} (2006) 004
  [arXiv:hep-ph/0601083]. 

 \bibitem{Nardi:2006fx}
  E.~Nardi, Y.~Nir, E.~Roulet and J.~Racker,
  JHEP {\bf 0601} (2006) 164
  [arXiv:hep-ph/0601084].

 \bibitem{aba06b}
  A.~Abada, S.~Davidson, A.~Ibarra, F.~X.~Josse-Michaux, 
M.~Losada and A.~Riotto,
  JHEP {\bf 0609}  (2006) 010
  arXiv:hep-ph/0605281.

\bibitem{ba00}
  R.~Barbieri, P.~Creminelli, A.~Strumia and N.~Tetradis,
  Nucl.\ Phys.\ B {\bf 575}, 61 (2000)
  (for the updated version of this paper see [arXiv:hep-ph/9911315]).

\bibitem{en03-fu05}
  T.~Endoh, T.~Morozumi and Z.~h.~Xiong,
  Prog.\ Theor.\ Phys.\  {\bf 111}, 123 (2004)
  [arXiv:hep-ph/0308276]; 
%
  T.~Fujihara, S.~Kaneko, S.~Kang, D.~Kimura, T.~Morozumi and M.~Tanimoto,
  Phys.\ Rev.\ D {\bf 72}, 016006 (2005)
  [arXiv:hep-ph/0505076].

 \bibitem{Davidson:2008bu}
  S.~Davidson, E.~Nardi and Y.~Nir,
 Phys.\ Rept.\  {\bf 466}, 105 (2008); arXiv:0802.2962 [hep-ph].

\bibitem{review}
  A.~Strumia,
  arXiv:hep-ph/0608347;
  E.~Nardi,
  arXiv:hep-ph/0702033;
  Y.~Nir,
  arXiv:hep-ph/0702199;
  M.~C.~Chen,
  arXiv:hep-ph/0703087;
   E.~Nardi,
   arXiv:0706.0487 [hep-ph];
  A.~Pilaftsis,
  arXiv:0904.1182 [hep-ph].

\bibitem{softlepto}
  Y.~Grossman, T.~Kashti, Y.~Nir and E.~Roulet,
  Phys.\ Rev.\ Lett.\  {\bf 91}, 251801 (2003)
  [arXiv:hep-ph/0307081];

\bibitem{D'Ambrosio:2003wy}
  G.~D'Ambrosio, G.~F.~Giudice and M.~Raidal,
  Phys.\ Lett.\  B {\bf 575}, 75 (2003)
  [arXiv:hep-ph/0308031];

\bibitem{Grossman:2004dz}
  Y.~Grossman, T.~Kashti, Y.~Nir and E.~Roulet,
  JHEP {\bf 0411}, 080 (2004)
  [arXiv:hep-ph/0407063].

 \bibitem{bu01} 
W.~Buchmuller and M.~Plumacher,
  Phys.\ Lett.\ B {\bf 511}, 74 (2001)
  [arXiv:hep-ph/0104189]. 

 \bibitem{na05}
  E.~Nardi, Y.~Nir, J.~Racker and E.~Roulet,
  JHEP {\bf 0601}, 068 (2006) [arXiv:hep-ph/0512052].



\bibitem{ha90}
  S.~Y.~Khlebnikov and M.~E.~Shaposhnikov,
  Nucl.\ Phys.\  B {\bf 308}, 885 (1988). J.~A.~Harvey and M.~S.~Turner,
  Phys.\ Rev.\ D {\bf 42}, 3344 (1990).
  H.~K. ~Dreiner and G. ~G. ~Ross, Nucl.\ Phys.\ B {\bf 410}, 188 (1993).

\bibitem{susy-equilibrium} 
  L.~E.~Ibanez and F.~Quevedo,
  Phys.\ Lett.\  B {\bf 283}, 261 (1992)
  [arXiv:hep-ph/9204205];
  T.~Inui, T.~Ichihara, Y.~Mimura and N.~Sakai,
  Phys.\ Lett.\  B {\bf 325}, 392 (1994)
  [arXiv:hep-ph/9310268].


\bibitem{eR-equilibrium}
  J.~M.~Cline, K.~Kainulainen and K.~A.~Olive,
  Phys.\ Rev.\ Lett.\  {\bf 71}, 2372 (1993)
  [arXiv:hep-ph/9304321];
  J.~M.~Cline, K.~Kainulainen and K.~A.~Olive,
  Phys.\ Rev.\  D {\bf 49}, 6394 (1994)
  [arXiv:hep-ph/9401208].

 \bibitem{Rubakov:1996vz}
  V.~A.~Rubakov and M.~E.~Shaposhnikov,
  Usp.\ Fiz.\ Nauk {\bf 166}, 493 (1996)
  [Phys.\ Usp.\  {\bf 39}, 461 (1996)]
  [arXiv:hep-ph/9603208];
  D.~J.~H.~Chung, B.~Garbrecht and S.~Tulin,
  JCAP {\bf 0903}, 008 (2009)
  [arXiv:0807.2283 [hep-ph]].

\bibitem{Engelhard:2006yg}
  G.~Engelhard, Y.~Grossman, E.~Nardi and Y.~Nir,
  Phys.\ Rev.\ Lett.\  {\bf 99}, 081802 (2007)
  [arXiv:hep-ph/0612187].

\bibitem{softflavor}
  C.~S.~Fong and M.~C.~Gonzalez-Garcia,
  JHEP {\bf 0806}, 076 (2008)
  [arXiv:0804.4471 [hep-ph]];
  C.~S.~Fong and M.~C.~Gonzalez-Garcia,
  JHEP {\bf 0903}, 073 (2009)
  [arXiv:0901.0008 [hep-ph]].

\bibitem{PFL}
  D.~Aristizabal Sierra, M.~Losada and E.~Nardi,
  Phys.\ Lett.\  B {\bf 659}, 328 (2008)
  [arXiv:0705.1489 [hep-ph]];
  D.~A.~Sierra, L.~A.~Munoz and E.~Nardi,
  arXiv:0904.3043 [hep-ph]; 
  arXiv:0904.3052 [hep-ph].

\end{thebibliography}
\end{document}